\documentclass[preprint]{aastex63}
\let\tablenum\relax
\usepackage{siunitx}
\usepackage{soul}
\usepackage{amsmath}
\usepackage{siunitx}
\usepackage{natbib}
\usepackage{epstopdf}
\usepackage{float}
\usepackage[caption = false]{subfig}
\usepackage{graphicx,array,multirow}
\usepackage{listings}
\graphicspath{{./}{figures/}}

\newcommand{\myeq}{\mathrel{\overset{\makebox[0pt]{\mbox{\normalfont\tiny\sffamily $\zeta=2$}}}{\longrightarrow}}}

\newcommand{\lapprox} {\, \lower3pt\hbox{$\sim$}\llap{\raise2pt\hbox{$<$}}\,}
\newcommand{\gapprox} {\, \lower3pt\hbox{$\sim$}\llap{\raise2pt\hbox{$>$}}\,}

\def\mt#1{{\color{black}{#1}}}

\begin{document}



\title{Relating the Solar Wind Turbulence Spectral Break at the Dissipation Range with an Upstream Spectral Bump at Planetary Bow Shocks}

\author[0000-0003-4747-6252]{M. Terres}
\affil{Department of Space Sciences, University of Alabama in Huntsville, USA, Michael.Terres@uah.edu}

\author[0000-0003-4695-8866]{Gang Li}

\affil{Department of Space Sciences, University of Alabama in Huntsville, USA, gangli.uah@gmail.com}

\begin{abstract}
At scales much larger than the ion inertial scale and the gyro-radius of thermal protons, magnetohydrodynamic (MHD) theory is well equipped to describe the nature of solar wind turbulence. The turbulent spectrum itself is defined by a power law manifesting the energy cascading process. A break in the turbulence spectrum develops near ion scales, signaling the onset of energy dissipation. The exact mechanism for the spectral break is still a matter of debate. In this work, we use the 20 Hz \textit{MESSENGER} magnetic field data during four planetary flybys at different heliocentric distances to examine the nature of the spectral break in the solar wind. 
We relate the spectral break frequencies of the solar wind MHD turbulence, \mt{found in the range of $0.3$ to $0.7$ Hz, with the well-known characteristic spectral bump at frequencies $\sim 1$ Hz upstream of planetary bow shocks. Spectral breaks and spectral bumps  during three planetary flybys are identified from the \textit{MESSENGER} observations, with heliocentric distances in the range of $0.3$ to $0.7$ au.  The \textit{MESSENGER} observations are complemented by one \textit{MMS} observation made at 1 au.}  {We find that the ratio of the spectral bump frequency to the spectral break frequency appears to be $r$- and $B$- independent. From this we postulate that the 
wave number of the spectral break and the frequency of the spectral bump have the same dependence on the magnetic field strength $|B|$.  The implication of our work on the nature of the break scale is discussed. }
\end{abstract}


\section{Introduction}
%
%

Turbulence in the solar wind consists of fluctuations across various scales. Nonlinear interactions cascade energy, through the inertial range, across several frequency (wavenumber) decades with a Kolmogorov $k^{-5/3}$ scaling \citep{Bruno2013}. Upon reaching ion kinetic scales, the power spectral density of magnetic field fluctuations develops a break, above which the turbulent cascade steepens 
\citep{Leamon1998b, Smith2006, Sahraoui2010, Alexandrova2013}.
\mt{At sub-ion scales, spectrum scaling exponent of $-7/3$ is predicted from gyrokinetic theory while numerical simulations have recovered a $-8/3$ exponent \citep{Schekochihin2009,Boldyrev2012}. Observations often show a $-2.8$ spectra, but steeper spectra have also been observed \citep{Sahraoui2010,Alexandrova2013}. At even smaller scales (near and beyond electron scales), a $f^{-5.5}$ spectrum from MMS observations has been reported by \cite{Macek2018} (see Figure 2 in \cite{Macek2018}). The observation of \citet{Macek2018} agrees with the predictions of $-16/3$ from kinetic theory \citep{Schekochihin2009}. Incidentally, we note that a $k^{-5.5}$ spectrum has been inferred to occur at solar flare site by \citet{Li2021_GRL}.
} 
The ion kinetic scale frequency is associated with various dissipation or dispersive mechanisms that lead to heating of the surrounding plasma. The differences between dispersive or dissipation mechanisms have been investigated recently \citep{Bowen2020}. 


Several scales have been suggested to cause the spectral break. The first is the ion inertial length scale, $\lambda_{i} = c/\omega_{pi}=V_A/\Omega_p$, and the second is the ion Larmor radius, 
{$\rho = v_{th, \perp}/\Omega_p$. Here $\omega_{pi} = \sqrt{(4\pi n_i q^2)/m_{i}}$ and $\Omega_{p} = q B/ m_{i} c$ are the ion plasma frequency and ion cyclotron frequency respectively with $c$ the speed of light and $v_{th,\perp}$ the ion perpendicular thermal speed.} The ion inertial length is associated with the thickness of current sheets.  Current sheets have long been recognized as a significant place for plasma heating through magnetic reconnection \citep{Leamon2000}. Besides, ion inertial length scale is also  related to the Hall effect, which represents another mechanism 
for efficient dissipation \citep{Bourouaine2012}. In contrast,  $\rho$ is associated with Landau damping of Alfv\'{e}n waves that heats the surrounding ion population \citep{Schekochihin2009}.  

The corresponding wave numbers for $\rho_i$  and $\lambda_{i}$ are represented as $k_L=1/\rho_i$ and $k_i=1/\lambda_{i} = k_L \beta_p^{1/2}$ respectively, where $\beta_p = (v_{th}/V_A)^2$ is the plasma beta. 
A third scale that has been related to the dissipation range spectral break is the cyclotron resonance scale, in which parallel propagating Alfv\'{e}n waves exchange energy with the surrounding thermal ion population through the cyclotron resonance $\omega_r + k_{||} v_{th,||} = \Omega_p$ \citep{Leamon1998,Smith2006,Bruno2014,Woodham2018}
The minimum resonant wavenumber is  $k_d = \Omega_p/(v_{A} + v_{th,||})  = k_L /(1+\beta^{-1/2}_{||})$. At 1 AU, the plasma beta of solar wind is 
often close to $1$, so the two scales $k_L^{-1}$ and $k_i^{-1}$ are close to
each other. To examine the effect of plasma $\beta$ on the spectral break locations, \citet{Chen2014} identified intervals that have extreme value of $\beta_{\perp}$ ($<<1$ and $>>1$ respectively) and studied the location of the spectral break.  These authors found that that for $\beta<<1$ the break occurs at 
$k_i$ and for $\beta>>1$, the break occurs at $k_L$. The finding of 
\citep{Chen2014} implies that the dissipation can undergo different processes, and as the solar wind MHD turbulence cascading proceeds, the process that has the largest characteristic scale prevails.



{Observations from various heliocentric distances were used to constrain the spectral break location.}
By examining how the spectral break location varies with heliocentric distances, \cite{Perri2010} and \citet{Bourouaine2012} suggested that the break location, invariant of distance, agreed with the ion inertial scale. However, a later study by \citet{Bruno2014} who made use of special radial alignments of three spacecraft in the solar wind (\textit{MESSENGER}, \textit{Wind}, and \textit{Ulysses}), showed that the break location agrees best with the ion cyclotron resonance scale as first predicted by \citet{Leamon1998}.  
{\cite{Telloni2015} investigated the high-frequency magnetic fluctuations beyond the spectral break, for the periods identified by \cite{Bruno2014}.}
They find that the characteristics are compatible with {both left-hand outward-propagating ion cyclotron waves and right-hand kinetic Alfv\'{e}n waves \citep{Telloni2015}. The existence of kinetic Alfv\'{e}n waves and ion cyclotron waves found near the break location implies that the dissipation can be associated with ion-cyclotron resonance and/or Landau damping \citep{Telloni2015}}.

In a more recent study, \cite{Woodham2018} using data from \textit{Wind} spacecraft, performed an extensive study of  the effect of $\beta_p$ on the break frequency location. They concluded that the proton cyclotron resonance scale is most closely related to the spectral break when $\beta_p \sim 1$, but no definitive conclusions could be made at extreme $\beta_p$ values.  
\cite{Wang2018} investigated the $\beta_p$ dependence of the break frequency for a broad range of $\beta_p$ from $0.005$ to $20$ and also found that cyclotron resonance best describes the spectral break location, although at extreme $\beta_p << 1$ and $\beta_p >> 1$, inertial length scale or proton gyroradius can not be ruled out \citep{Wang2018}. In a recent work, \citet{Duan2020} used data from the \textit{Parker Solar Probe} (\textit{PSP}) to examine the radial evolution of ion spectral breaks during the second solar encounter, and obtained results that support \cite{Woodham2018, Wang2018}. 

{A fourth scale, the disruption scale $\lambda_D = C_D L_{\perp}^{1/9} (d_e \rho_s )^{4/9}$, has been proposed by \citet{Mallet2017a} and \citet{Loureiro2017}. The expression for $\lambda_D$ differ in 
 \citep{Mallet2017a} and \citep{Loureiro2017} but are close.  
Here $L_{\perp}$ is the outer scale of the inertial range, $d_e = c/\omega_e$ is electron inertial length, and $\rho_s=\rho \sqrt{ZT_e/2T_i} \sim \rho$ where $\rho$ is the electron gyroradius. The disruption scale occurs when current sheets are destroyed by magnetic reconnection. 
\mt{It has been pointed out that turbulence and magnetic reconnection are intimately related \citep{Sundkvist2007, Karimabadi2014}. Reconnection occurs when electrons cannot supply the current needed to support antiparallel magnetic fields. Consequently ions and electrons decouple and at smaller scales electron physics dominates \citep{Macek2019}. At these scales a steeper spectrum has been reported by \citet{Macek2018} using MMS observations in the Earth's magnetosheath. 
Behind the Earth's bow shock, magnetic reconnection occurs frequently in the magnetosheath than in the pristine solar wind \citep{Karimabadi2014}, making the magnetosheath an ideal place to investigate the turbulence spectrum at the electron and sub-electron scales.} 

Using over 12 years of Wind data, \citet{Vech2018} compared the 
turbulence spectral break location with the above four scales, $\rho$, $\lambda_i$,  {$\lambda_i+\sigma$ (where $\sigma = v_{th_{\parallel}/\Omega_p}$)}, and $\lambda_D$. They found that the ratio of the spectral break location to {the disruption and cyclotron resonance scale} (after converting to the frequency regime assuming Taylor hypothesis) are relatively constant for a broad range of $\beta$, implying that these two scales may be the underlying causes of the spectral break. For a subset of the data where the dissipation range shows a steeper (than $k^{-3}$) spectrum, due to magnetic reconnection, they further find that the disruption scale is in a better agreement with the break location than the cyclotron resonance scale.}

While the above studies focus on the power spectra break of the quiet solar wind, another well-known and unique spectral feature, a spectral bump, has been first reported by \citet{Fairfield1969} using observations upstream of Earth's bow shock. It has subsequently been observed upstream at multiple planetary bow shocks and has been referred to as the ``1 Hz" wave in the literature \citep{Brain2002, Le2013, Wilson2016,Xiao2020}. {''1 Hz" waves are characterized, in the plasma rest frame, as small amplitude right-hand polarized whistler waves generated upstream of collisionless shocks (i.e. planetary bow shocks) \citep{Le2013, Wilson2016}.} The bumps associated with these waves occur at frequencies ($\sim 1$ Hz) higher than the spectral breaks, in the range of $20-100 \Omega_{pi}$ \citep{Hoppe1981}. The detailed generation mechanism of these waves, however, is still unknown. Nevertheless, it has been noted by \citep{Russell2007} that the bump locations is proportional to magnetic field $|B|$. 
The \textit{Mercury Surface, Space Environment, Geochemistry, and Ranging} (\textit{MESSENGER}) spacecraft was launched in 2008.  It has a few  planetary flybys at both Venus and Mercury during which it traversed their bow shocks. During  these flybys, MESSENGER was able to observe both the spectral break and the spectral bumps within a short period of time. As we show here, this provides a rare opportunity to examine the nature of the spectral break by using the spectral bump as a marker. 
 
{To better resolve spectral features, especially at frequencies close and higher than the spectral break,  we adopt the wavelet transforms to examine the power spectral density (PSD). Wavelet analysis has become an increasingly popular tool to study a signals PSD due to the localization of the wavelet components in both frequency and time \citep{Horbury2008,Podesta2009,Woodham2018}. Unlike the traditional Fourier transform, wavelet analysis produces smoother spectral plots, allowing spectral features to be better identified. 
}

{In this manuscript, we analyze the spectral break and spectral bump for $3$ planet flybys, one at Venus, and two at Mercury using MESSENGER. We complement the MESSENGER periods by two periods from the Magnetospheric Multiscale (MMS) spacecraft at the Earth, covering a heliocentric distance from $0.34$ to $1$ au. The power spectral density is computed by wavelet analysis. Spectral breaks and spectral bumps are identified. We compare the location of these two spectral features and find that the ratio of these two frequencies are largely independent of $|B|$ and $r$. We discuss the implication of our results and the impact it has on the dissipation scale.}


\section{Data and Methods}
We use magnetic field data from the MESSENGER/MAG instrument \citep{Anderson2007} for one flyby at Venus and two 
flybys at Mercury. For each flyby, two periods, one in the dawn side and one in the dusk side, are identified. 
During planetary flybys, \textit{MESSENGER/MAG} recorded magnetic field data at $20$ Hz. \mt{Since the 
spectral breaks and spectral bumps discussed in this work are below or at $\sim$ 1 Hz, so $20$ Hz resolution is} sufficient for our study of ion scales \cite{Bruno2014}.
However, as noted by \citet{Anderson2007}, there are possible digitization effects above $3$ Hz, leading to some instrument noises.
As we discuss below, however, except one spectral bump during a dusk period,  the spectral breaks and bumps from MESSENGER/MAG observations occur below $3$ Hz. We also use observations at 1 AU from the  dual fluxgate magnetometers \citep{Russell2016}
onboard the Magnetospheric Multiscale (\textit{MMS}) spacecraft. {\textit{MMS} provides magnetic field data at 16 Hz resolution \citep{Russell2007}. Overall, since analyses are made from a single spacecraft, we negate any errors in the spacecraft measurements.}

\begin{figure}[htb]
    \centering
    \includegraphics[height=6.5cm]{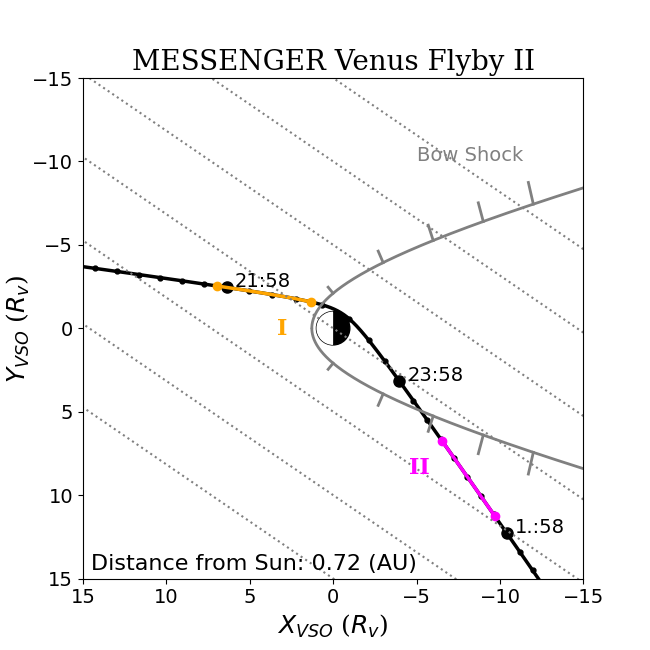}
    \includegraphics[height=6.5cm]{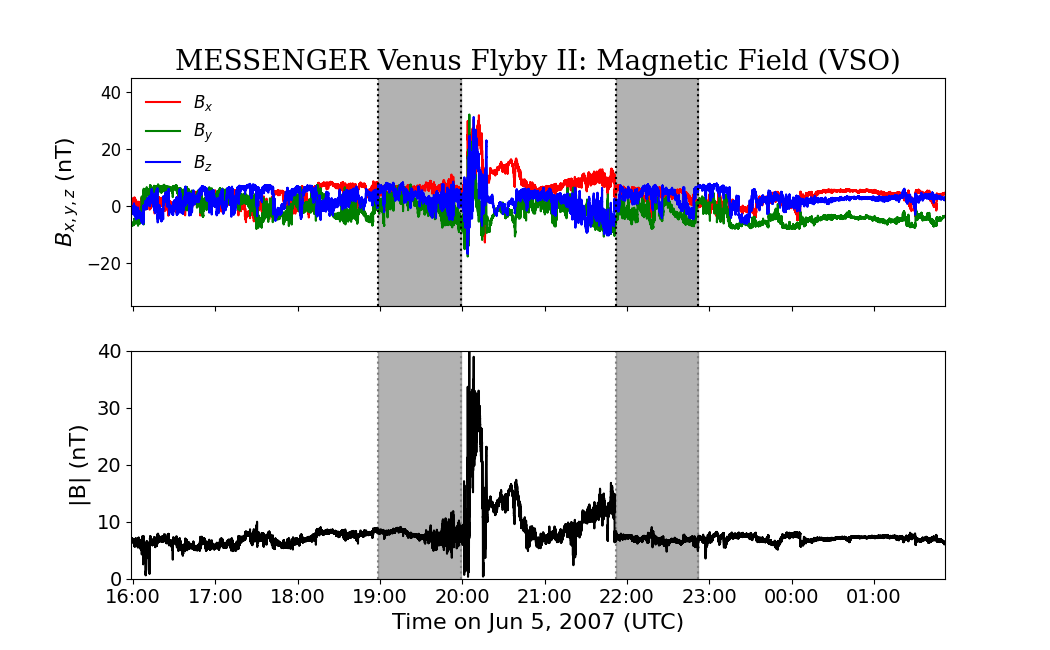}
    \medskip
    \includegraphics[height=6.5cm]{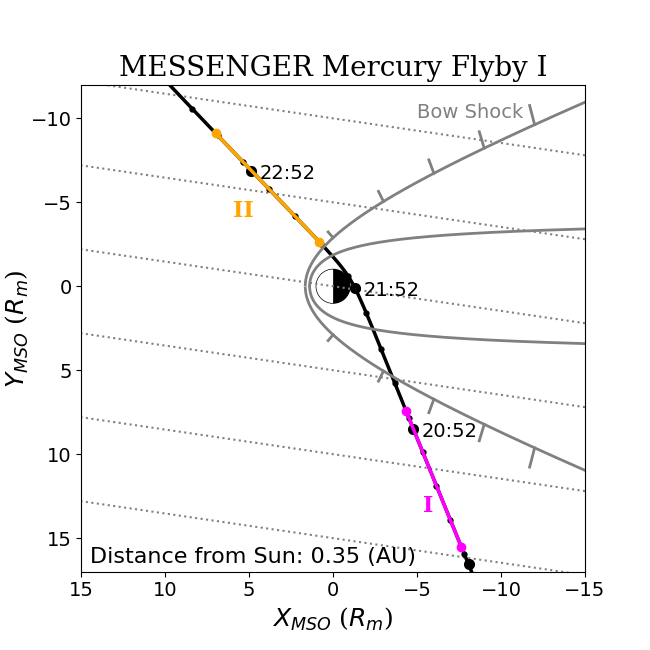}
    \includegraphics[height=6.5cm]{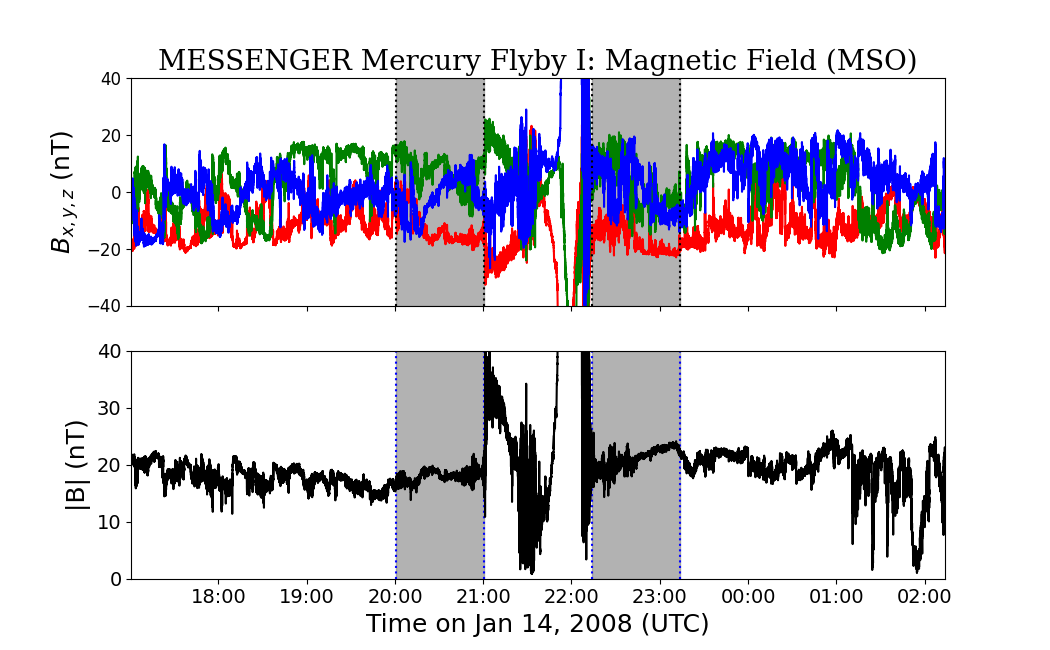}
    \medskip
    \includegraphics[height=6.5cm]{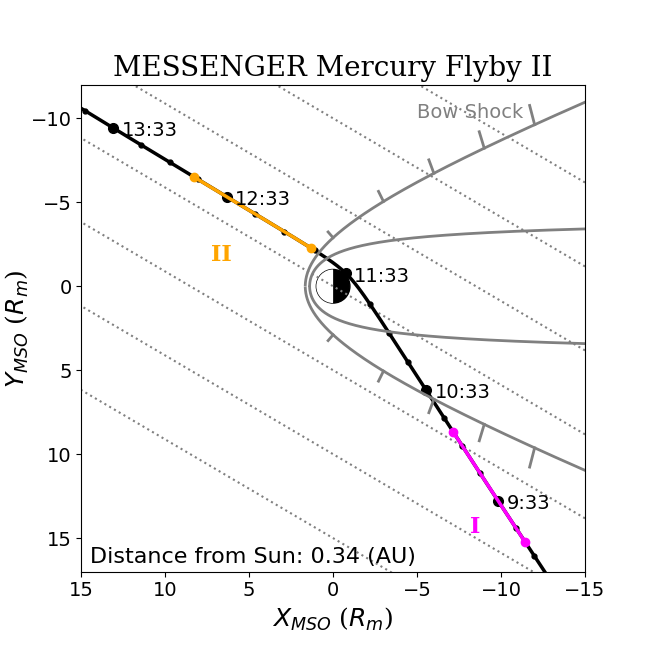}
    \includegraphics[height=6.5cm]{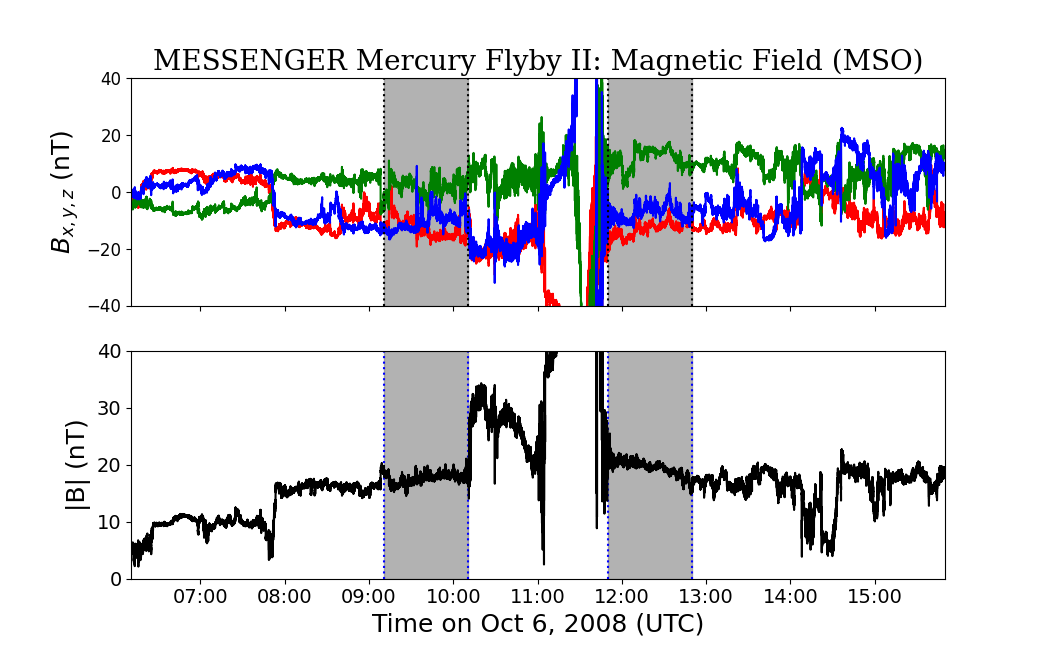}
    \caption{\textit{MESSENGER} trajectories and magnetic fields during planetary flybys of Venus and Mercury. Left panels show the spacecraft trajectories during each planetary flyby \mt{(black)}. The \mt{magenta} (orange) segments indicate periods of enhanced upstream wave activities ins the dusk (dawn) side. The direction of spacecraft trajectory is shown in the order of I and II. Gray curves show the bow shock, using approximation form \citet{Slavin1984}. The right panels plot the corresponding magnetic field components and magnitude. Greyed periods correspond to the \mt{magenta} and orange segments in the left panels. }
    \label{path_bfield}
\end{figure}

Figure \ref{path_bfield} shows the trajectories of the three planetary flybys (6 periods) and the corresponding magnetic field measurements. 
Each period is four hours in duration. A one-hour sub-period, in each period, close to the bow shock, is colored in \mt{magenta} for 
the dusk side and in orange for the dawn side (we adopt this color code convention throughout this work). 
The same one-hour period is shown as the shaded area on the right panels of Figure \ref{path_bfield}. Within this one-hour sub-period, the spacecraft is closer to the bow shock and the upstream ``1 Hz" waves are observed (yielding a spectral bump). The remaining $3$-hour sub-period is further away from the bow shock and no upstream ``1 Hz" waves are observed (yielding no spectral bump). Each period is also labeled with  ``I" or ``II" to denote inbound or outbound trajectory. 

\begin{figure}[htb]
    \centering
    \includegraphics[height = 6.5cm]{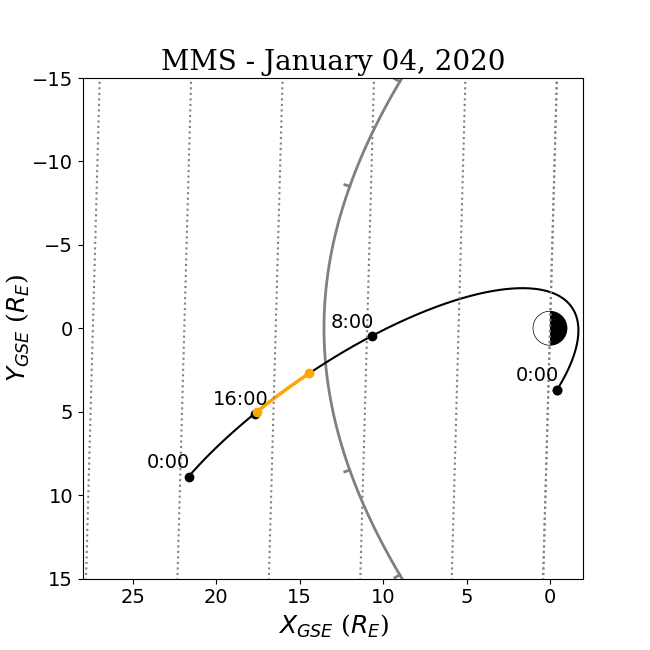}
    \includegraphics[height=6.5cm]{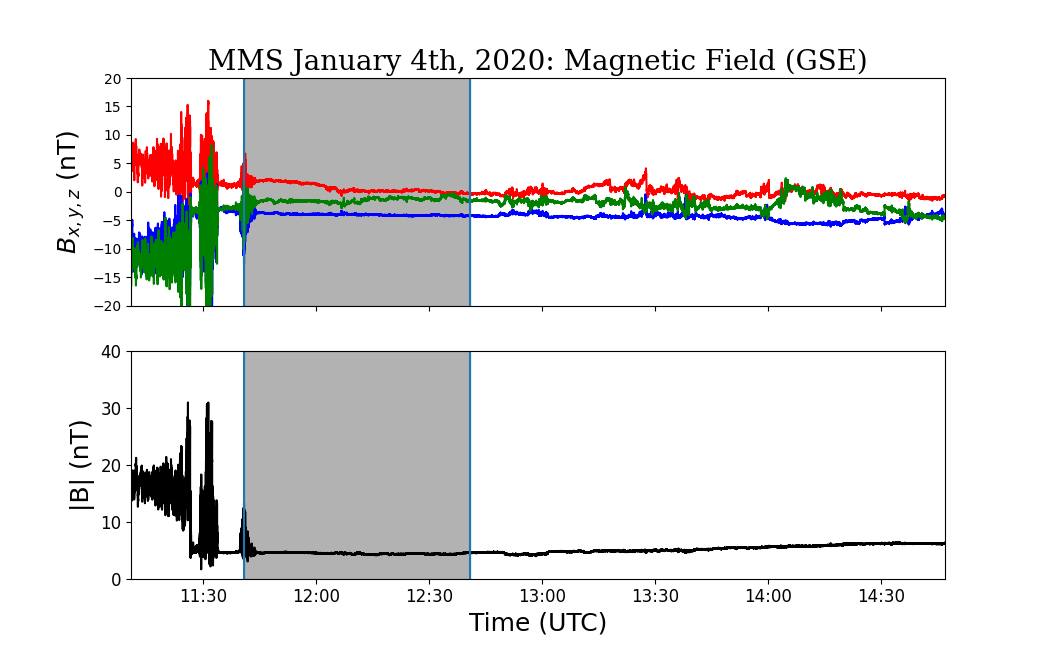}
    \medskip
    \includegraphics[height=6.5cm]{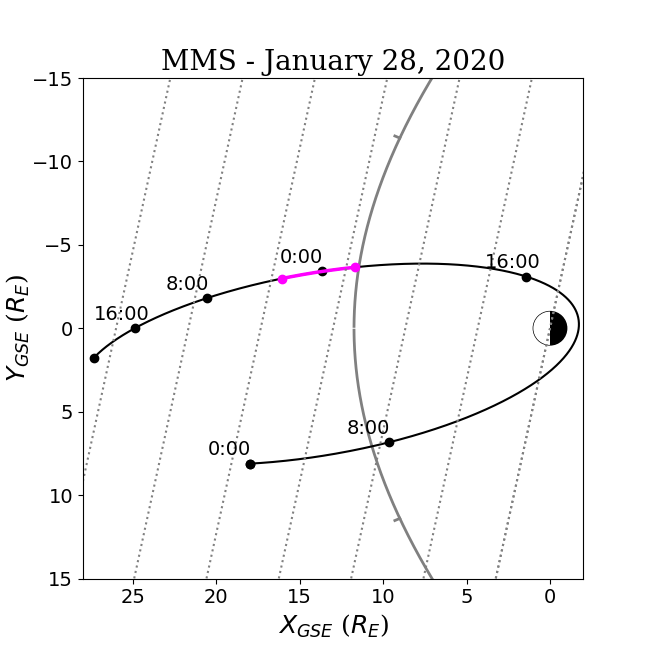}
    \includegraphics[height=6.5cm]{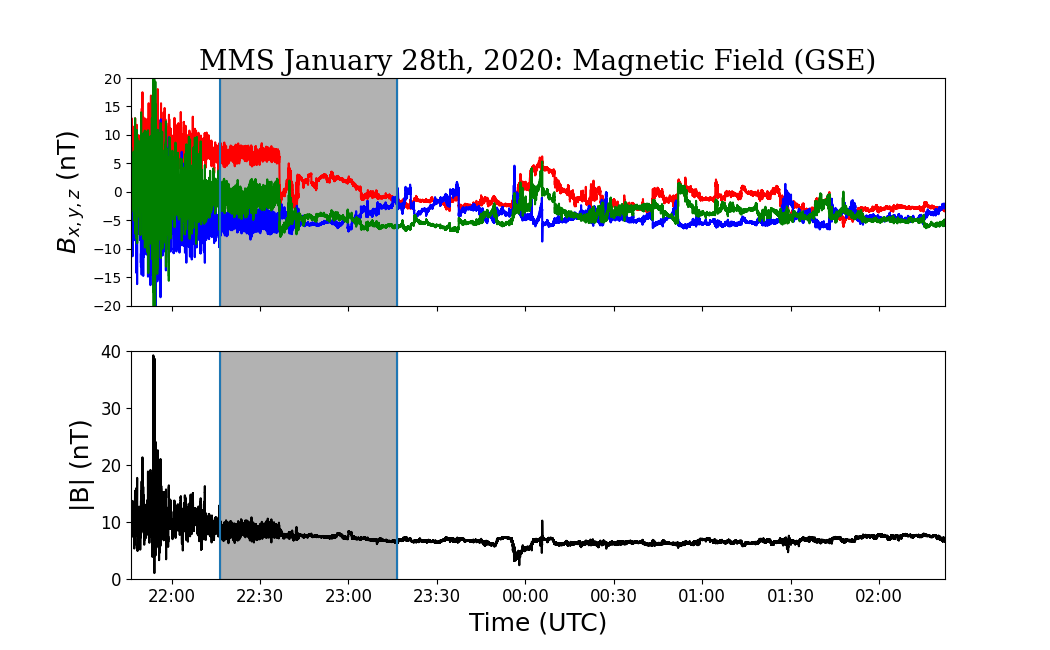}
    \caption{Similar to Figure~\ref{path_bfield} but for \textit{MMS} observations. The left figures show \textit{MMS} 1 spacecraft position during the two selected orbits. The regions marked in \mt{magenta and orange} correspond to the four hour duration post bow shock crossing. The panels to the right correspond to the magnetic field components and magnitudes. The greyed regions correspond the periods of enhanced wave activity. 
    {Note that the bow shock location for the January 28 event is shifted by $1.8$ Re from the model estimate since the magnetic field measurement for the period labeled by \mt{magenta} is outside the bow shock. }
    }
    \label{MMS_Bfield}
\end{figure}

The top panel is for June 5th, 2007, during the second Venus flyby. 
The middle and bottom panels are for January 14th and October 6th, 2008, during the first and second Mercury flybys, respectively. 
The bow shock positions shown in the left panels of Figures~\ref{path_bfield} and \ref{MMS_Bfield} are obtained using the functional form described \mt{by} \cite{Slavin1984}, and the dashed grey lines  depict the average background magnetic field direction for our selected solar wind periods.

Similar to Figure~\ref{path_bfield}, Figure~\ref{MMS_Bfield} shows two \textit{MMS} trajectories and magnetic field data for January 4th and 28th, 2020. 
The magnetic field data used was during the fast mode survey which has a resolution of $16$ Hz. \mt{
MMS also has a BURST mode which would allow one to study breaks at even higher frequencies and steeper slopes \citep{Macek2018}.}
Unlike the trajectories of MESSENGER Venus/Mercury flybys, it is hard to denote ``dawn" or ``dusk" to these two MMS periods. However, the trajectory on the January 28th 2020 (hereafter TR1) is more along the Earth-Sun direction than that (hereafter TR2) of the  January 4th 2020. Consequently we expect the 3-hour period from which we obtain the spectral break in TR1 is of more 
pristine solar wind than that in the corresponding 3-hour period in TR2.  In the following we denote the trajectory on 2020 January 4 as``dusk side" and that on 2020 January 28 as ``dawn side" in the following analysis. 
\mt{In Figure~\ref{MMS_Bfield}, we use the same color convention, i.e., magenta denotes dusk and orange denotes dawn.}

%
%




To calculate the power spectral density (PSD), we follow  \cite{Podesta2009} using a wavelet analysis.
Comparing to traditional spectral analysis methods based on Fourier transform,
wavelet analysis yield smoother spectra, allowing clear identification of spectral features. 
Note that plasma data from \textit{MESSENGER} \mt{are} not available due to the sunshade restricting the Fast Imaging Plasma Spectrometer (FIPS) field of view \citep{Raines2011}. 
Previous investigations around Mercury have approximated solar wind conditions using the Ulysses spacecraft \citep{Korth2011}. 
\cite{Baker2011} estimated solar wind plasma properties at the three Mercury flybys using the WSA-Enlil model.
Our result in this paper does not require knowledge of solar wind plasma property, although knowing solar wind speed helps to 
interpret our results. To obtain a rough estimate of the solar wind speed, we assume the solar wind can be approximated as quasi-steady structures and shift the \textit{ACE} measurement to locations at the \textit{MESSENGER} by co-rotation. 




\section{Results and Discussion}

\begin{figure}[h]
    \centering
    A)
    \includegraphics[width=0.36\linewidth]{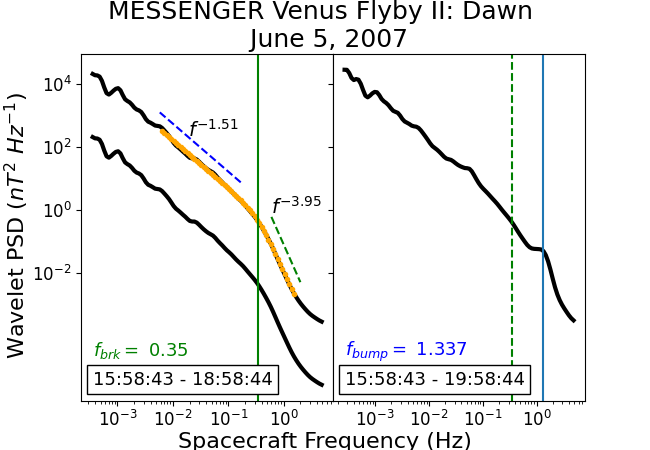}
    \centering
    B)
    \includegraphics[width=0.36\linewidth]{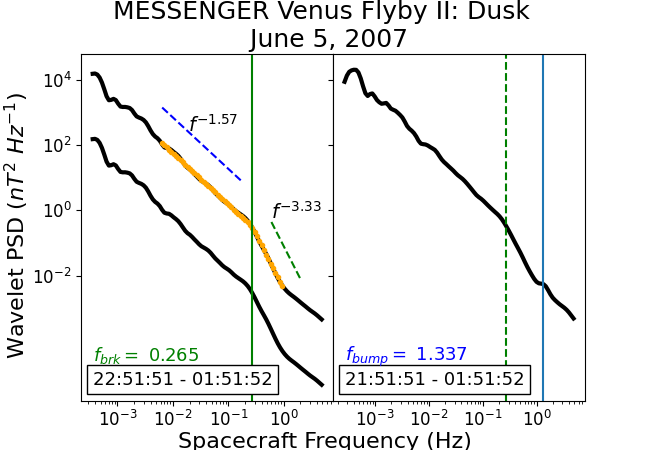}
    
    \centering
    {C)}
    \includegraphics[width=0.36\linewidth]{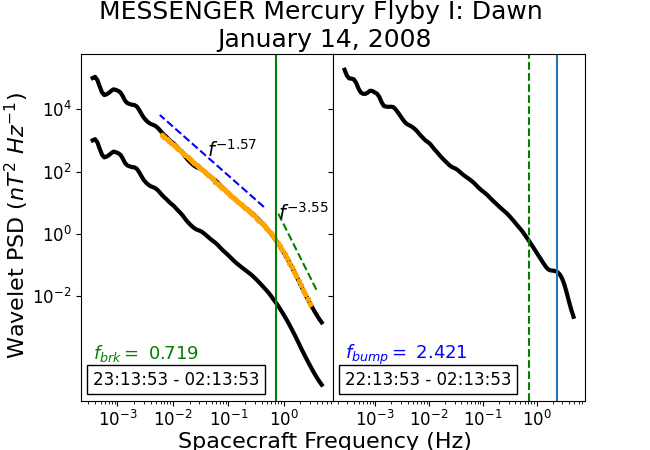}
    \centering
    {D)}
    \includegraphics[width=0.36\linewidth]{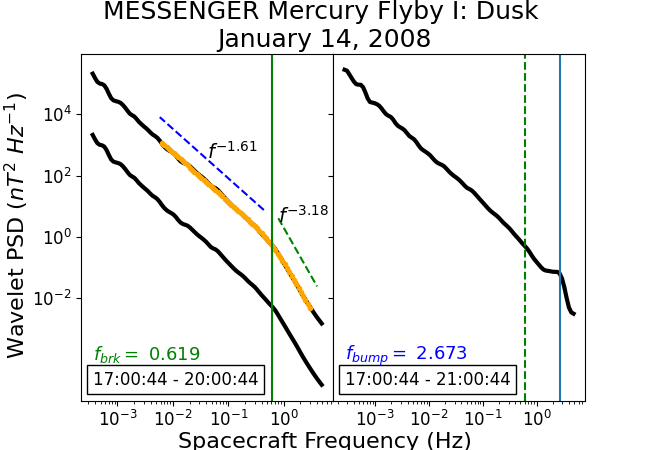}
    
    \centering
    {E)}
    \includegraphics[width=0.36\linewidth]{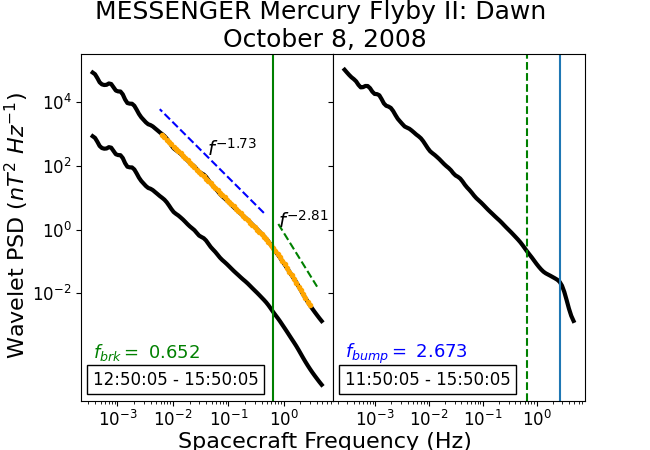}
    \centering
    {F)}
    \includegraphics[width=0.36\linewidth]{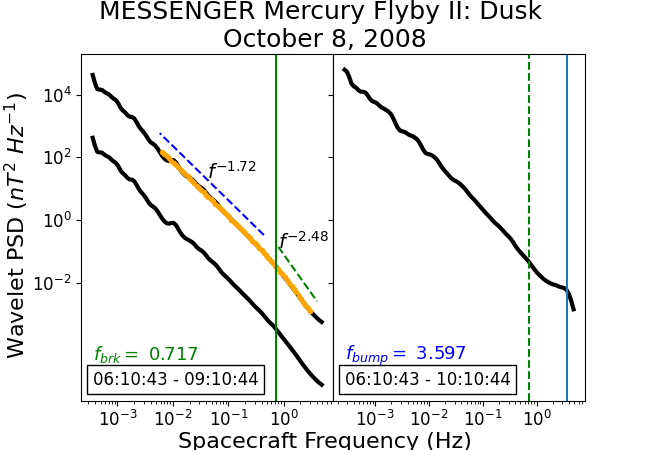}
    
    \centering
    {G)}
    \includegraphics[width=0.36\linewidth]{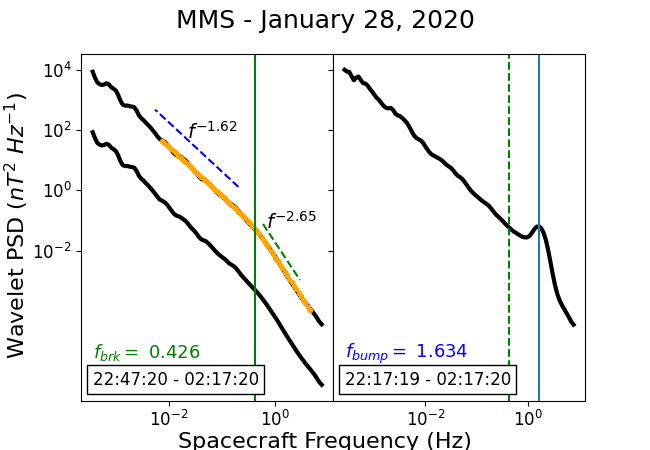}
    \centering
    {H)}
    \includegraphics[width=0.36\linewidth]{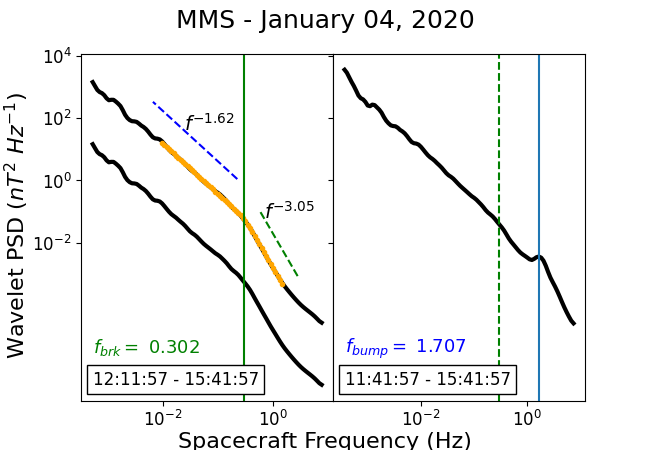}
    \caption{ Power spectra for periods shown in Figure~1 and Figure~2.
    In each panel, the left side shows the $3$-hour solar wind power spectra and the right side shows the full $4$-hour solar wind power spectra. Significant upstream wave activities are only present close to the bow shock and result in a clear bump around $1$ Hz for the $4$-hour power spectra.  
    The blue (green) dashed line is the fitting for the inertial (dissipation) range. The vertical solid green line indicates the frequency of the spectral break. The vertical blue solid line indicates the frequencies of the bump ``peak". The vertical dashed green line marks the same break frequency  
    as the solid green line from the left side. The x-axis is the spacecraft frequency and the y-axis is the power density.}
    \label{Wavelet_MESSENGER}
\end{figure}

We first calculate the power spectral density (PSD) for the eight periods we identify. 
Figure \ref{Wavelet_MESSENGER} shows the calculated wavelet spectrum for the $8$ periods in Figure \ref{path_bfield} and \ref{MMS_Bfield}.
The left panel shows the spectrum for the 3-hour period excluding the grey shaded areas in  
\ref{path_bfield} and \ref{MMS_Bfield}.
The trace of the wavelet spectrum, shown as the black curve, is plotted in the spacecraft frame, and shifted down by a 
factor of $100$. The top plot shows the trace in black and the corresponding fitting in orange.
We fit the spectrum using a broken power-law \citep{Liu2020}, given by,  
\begin{equation}
    p(f) \propto \frac{f^{\alpha}}{\big(1 + \big({f/f_{brk}}\big)^{\gamma}\big)^{\beta}}.
    \label{eq:fit}
\end{equation}
\mt{Equation~(\ref{eq:fit}) nicely fits a broken power law. At low frequencies, $p \sim f^{\alpha}$. When the frequency increases and becomes larger than $f_{brk}$,  $p$ transitions  to a steeper power law  $p \sim f^{\alpha - \beta \gamma}$. Parameters $\gamma$ and $\beta$ determine the spectral shape through the transition region.} To perform the fitting we first find the inertial range spectrum using a least-squares procedure. We fit the inertial range for Mercury from $5.8 \times 10^{-3}$ Hz to $0.44$ Hz and Venus and Earth from $5.8 \times 10^{-3}$ Hz to $0.17$ Hz. We reduce the fitting range to ensure all fits are before the break location. Using the least-squares method, we set the value of $\alpha$ in equation~(\ref{eq:fit}). We next identify the best-fit parameters for the break location $f_{brk}$ and spectral terms $\beta$ and $\gamma$. We calculate the dissipation spectrum by taking $\alpha - \beta\gamma$. To determine the best fit, we used a $\chi^{2}$ fitting routine, with the standard deviation being proportional to the PSD \citep{Liu2020}.   
To identify the bump location we calculated the local maxima in the frequency range from roughly 3 Hz to $f_{brk}$. 

To guide the eyes, \mt{in Figure \ref{Wavelet_MESSENGER}}, the blue dashed lines indicate the power law in the inertial range, and the dashed green lines indicate the power law in the dissipation range. The vertical solid green lines indicate the locations where the spectrum breaks. The right panel shows the spectrum for the full 4-hour period, including the grey shaded areas in Figure~\ref{path_bfield} and \ref{MMS_Bfield}. The dashed green lines in the right panel correspond to the break locations identified from the left panels (the solid green lines). The vertical blue lines indicate the bump ``peak" locations which are defined as the frequency $f_i$ at which  
the 2-point slope of the log of power $P(f)$ drops the most, i.e.,   
$(P^2(f_i)/(P(f_{i+1})*P(f_{i-1}))$ is the maximum at the spectral bump frequency.




Figure~\ref{Wavelet_MESSENGER} shows that the spectral bumps are due to wave activity occurring close to the bow shock, since they do not show up in the 3-hour period analysis and only show up in the 4-hour period. 
Furthermore, the spectral breaks are clearly seen for all the 3-hour data.  
For period ``D", corresponding to the ``dusk" side of the Mercury flyby II, the fitting yield a dissipation spectral index of $-2.48$, which is shallower than all other periods. A spectral break in this case is less clear than other periods.

One of our main findings is the correlation between the spectral break frequency $f_{brk}$ and the location of the spectral bump $f_{bump}$. {Since the former concerns the dissipation process at ion kinetic scale in the solar wind and the latter is related to whistler-mode waves that contains electron dynamics, the correlation between them is unexpected.} Note that both the spectral break and the spectral bump occur in the same flyby, hence instrument uncertainties are mitigated.



\begin{figure}[htb]
    \centering
    \includegraphics[width = 0.48\linewidth]{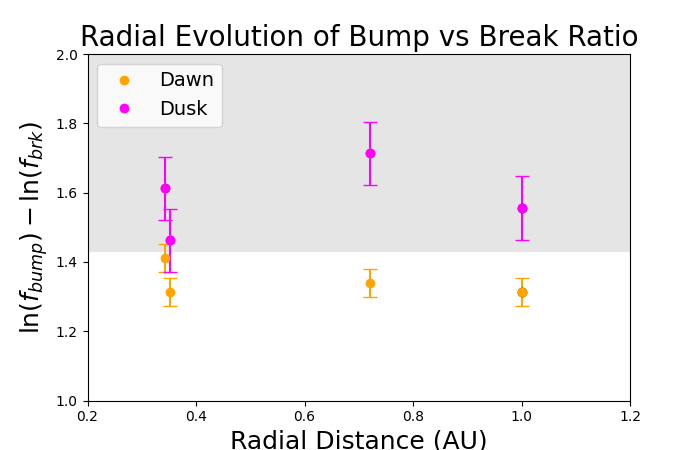}
    \includegraphics[width=0.48\linewidth]{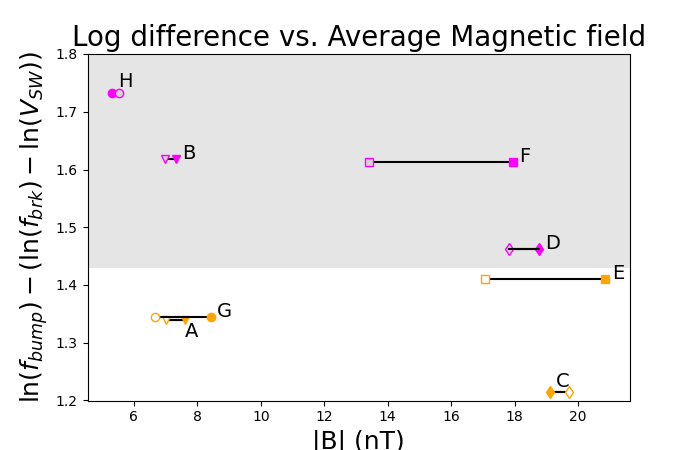}
    \caption{ The quantity $\eta$ as a function of $r$ (left) and $|B|$ (right).  
    The measurements are color coded with dawn events shown in orange and dusk events in \mt{magenta}. Data shows a clear separation between dawn side events and dusk side events. The error bars represent the standard error by treating dawn side events and dusk side events separately. In the right panel, two $B$ values are associated with each period.  The solid symbol is a 5-minute average when the upstream wave activity is clearly seen and the outlined value is a $3$-hour average in the solar wind. These $|B|$ values are listed in Table~1.  
    }
    \label{Radial_Evolution_2}
\end{figure}

We define a dimensionless parameter $\eta$ as $\eta = \ln(f_{bump}) - \ln(f_{brk})$, which is the log of the ratio $f_{bump}/f_{brk}$.  
The left panel of Figure~\ref{Radial_Evolution_2} shows $\eta$ as a function of radial distance, and the right panel shows $\eta$ as a function of $|B|$. 
We distinguish the dawn and dusk observations using orange (dawn) and \mt{magenta} (dusk) symbols.  We see a clear separation of $\eta$ between the dawn side trajectories and those in the dusk side. 
One possible explanation is that the solar wind turbulence is affected by wave activity originated from the bow shock. 
Since the solar wind propagates from the dawn side to the dusk side, the dawn side sees a more pristine solar wind. Note that although the upstream waves are seen clearly as a bump for the 4-hour period and not shown in the 3-hour period, these waves can propagate obliquely to the upstream and interact with the solar wind turbulence at a lower frequency.  This process will primarily 
affect the solar wind turbulence spectrum at the dusk side. 
As explained earlier, we categorize the MMS event on January 4th, 2020 as a dusk event, and that on January 28th, 2020 as a dawn event. \mt{Note that the break frequencies, as identified from Figure~\ref{Wavelet_MESSENGER}, are all in the range of $0.3$ to $0.7$ Hz. Furthermore
the MMS measurements have the smallest value, indicating that the break frequency decreases with increasing heliocentric distance.}
\textcolor{black}{As noted by \citet{Xiao2020}, the distribution and characteristics of these upstream waves can differ in the quasi-parallel and quasi-perpendicular regions of the bow shock. This is perhaps not surprising as one expects that the relative configuration between the solar wind magnetic field and the bow shock may play a role in the generation of these upstream waves.}


\begin{table}[h]
\caption{The $f_{bump}$, $|B|$, and $V^*_{sw}$ for the $8$ periods in Figure \ref{Wavelet_MESSENGER}.}
    \begin{tabular}{c c c c c c c c c} \hline
        Period: 
        &  A & B & C & D & E & F & G & H \\ \hline \hline
        $f_{bump}$ (Hz): 
        & 1.337  & 1.337  & 2.421  & 2.673  & 2.673  & 3.597  & 1.634  & 1.707  \\
        $|B|_{wave}$ (nT):
        \footnote{These are $5$-minute magnetic field when the upstream waves are clearly seen, except the MMS events where we consider only the first 2.4-minutes.} 
        & 7.61  & 7.32  & 19.10  & 18.77  & 20.86  & 17.94  & {8.416} & {5.314}  \\
        $|B|_{SW}$ (nT):
        \footnote{These are average magnetic field for the  $3$-hour solar wind period.}
        & 7.02  & 6.98  & 19.73  & 17.83  & 17.07  & 13.40  & 6.65  & 5.53   \\
        $V_{sw}$** (km/s):
        \footnote{These are estimates of solar wind speeds from \cite{Baker2011} using ENLIL simulations for the Mercury flybys.}
        & - & - & 420.0 & - & 380.0 & - & -  & -  \\
        $V_{sw}$* (km/s):
        \footnote{For A to F, these are estimates of solar wind speeds using ACE data. G and H are direct MMS measurements. The value for period H is from a later period due to lack of data availability in period H.}
        & 408.82 & 413.29 & 497.98 & 529.06 & 304.63 & 307.92 & 369.59  & 401.27   
    \end{tabular}    
    \label{tab:my_label}
\end{table}

Below we focus on the dawn side observations. What is interesting from Figure~\ref{Wavelet_MESSENGER} is that $\eta$ is almost a constant as a function of $r$ or $|B|$. Indeed, the range of $\eta$ is from {$1.21$} to {$1.41$}, translating to {$f_{bump}/f_{brk} = 3.35$ to $4.1$},  a factor of {$1.22$}. In comparison, the $|B|$ changes from $7.4$nT to $20.86$nT, a factor of $2.6$.  {This finding, that $\eta$, or $f_{bump}/f_{brk}$, is almost independent of $r$ or $|B|$, is our most important result.}
The variation of $\eta$ may in part due to the variation of $V_{sw}$ in these periods. If we assume the higher frequency $f_{bump}$ is of whistler wave in nature and only weakly depends on $V_{sw}$, then because  $\ln (f_{brk}) = \ln (k_{brk}) + ln (V_{SW})$ so $f_{brk}$ and therefore $\eta$ are also $V_{sw}$-dependent. Unfortunately, there was no plasma data from MESSENGER. One can nevertheless estimate $V_{sw}$ from either 1 AU measurements by ACE by taking into solar co-rotation, or use solar wind simulations to provide a crude estimate.
{From the right panel of Figure~4, we see that the variation of $\eta$
is largely due to the two Mercury flybys, periods E and C. 
Table I contains two different estimates of the solar wind speeds at Mercury fly-bys. 
The third row are the ENLIL model estimates for the two Mercury fly-bys from \cite{Baker2011}.
The fourth row are estimates using a ballistic solar wind plasma with a constant co-rotation assumption. We use 2-hour solar wind $V_{sw}$ measurements from ACE and backtrack it to the solar surface to find its footpoint and assume that the footpoint undergoes a  co-rotation with $\Omega_{\sun} = 1.54 \times 10^{-4}$ deg/s
to find the solar wind speed values at Mercury.   In both estimates, we see that the solar wind speed in period E is smaller than the other periods. In the ballistic estimate the solar wind speed in period C is also significantly larger than other periods. Consequently if we examine  the difference of $\ln (f_{bump}) - \ln (k_{brk})$, we find that it has a smaller range when compared to $\eta$ if using the estimates from \citep{Baker2011}, and a larger range if using the ballistic estimates.
It is possible that $\ln (f_{bump}) - \ln (k_{brk})$,  rather than $\ln (f_{bump}) - \ln (f_{brk})$, is $B$ or $r$ independent. 
} 

In the following discussions on the mechanism of $f_{brk}$, we 
assume that  $k_{brk}$ and $f_{bump}$ have the same $|B|$ and $r$ dependence.




We now discuss the implication of 
the very weak $|B|$ dependence of $f_{bump}/k_{brk}$ on the mechanism of the spectral break. First we note that $f_{bump}$  depends strongly on 
$|B|$. Earlier, \citet{Russell2007} suggested that $f_{bump}$ has a linear dependence on $|B|$. However, as shown in Table I, this dependence is nonlinear. Indeed, $f_{bump}$ may depend on other plasma parameters such as 
electron temperature $T_e$ and electron number density $n_e$, which themselves are correlated with $B$.
Unfortunately, MESSENGER has no plasma data available, preventing us from examining the explicit dependence of $f_{bump}$ on $T_e$ and $n_e$. 
Relating $f_{bump}$ to the $|B|$ as, 
\begin{equation}
    f_{bump} = C_0 \cdot \Omega_p^{\delta_1}, 
\label{eq:fbump}
\end{equation}
{where $C_0$ is a constant, then from the $|B|_{wave}$ and $f_{bump}$ in Table I for the dawn observations, we find $\delta_1 = 0.605 \pm 0.08$  with a correlation coefficient of $0.983$ using a least-squares fitting.}  
{Since we assume $\ln (f_{bump}) - \ln (k_{brk})$ is 
independent of $B$,  so we can write, 
\begin{equation}
    k_{brk} = C_0' \cdot \Omega_p^{\delta_1}, 
\label{eq:kbrk}
\end{equation}
where $C_0'$ is another constant \mt{and now $k_{brk}$ becomes a function of $\Omega_{p}$ only}. 
We now examine how the four possible scales: the ion inertial scale $k_i$, the ion gyration scale $k_L$, the cyclotron resonance scale $k_d$, and the disruption scale $k_D$, agree with equation~(\ref{eq:kbrk}).  
Because these measurements are made at multiple radial distances and at different times, we rewrite $k_i$, $k_L$, $k_d$, and $k_D$, 
using the following radial scalings: $n_p (n_e) \sim r^{-2}$,  $T \sim r ^{-0.66}$, and $B \sim r^{-\zeta}$. Inside $0.7$ au,  the radial component of the $\vec{B}$ dominates the transverse components and $\zeta \approx 2$; between $0.7$ au and to $1$ au, the transverse components of $\vec{B}$ becomes comparable to the radial component and we find $\zeta \approx -1.6$. The radial dependence of $T_p$ has been examined by \citet{Adhikari2020} using recent PSP observations.}

{First consider the ion inertial scale  $k_i = \Omega_{p}/V_{A}$. If the break is controlled by $k_i$ then it is independent of $B$. Therefore if $f_{bump}$ explicitly depends on $B$, then the fact that $\ln (f_{bump}) - \ln (k_{brk})$ is independent of $|B|$ can be used to argue strongly against $k_i$ as the underlying length scale for the dissipation spectral break. Of course, the radial dependence of $k_i$ through $n_i$ may lead to an apparent $B$ dependence which can be similar to that of $f_{bump}$. To examine how the radial dependence of $k_i$ can assume a $|B|$ dependence, we write,
\begin{equation}
    k_i/C_1 = (n_p)^{1/2} =  (n_0)^{1/2} \bigg(\frac{B}{B_0}\bigg)^{1/\zeta} =  \bigg(\frac{n_0}{B_0^{2/\zeta}}\bigg)^{1/2} B^{1/\zeta} \myeq
    \bigg(\frac{n_0}{B_0}\bigg)^{1/2} B^{1/2}, 
\label{eq:ki_cst}
\end{equation}
where $C_1$ is a constant, and $n_0$ and $B_0$ are the proton density and magnetic field when the plasma parcel is close to the Sun, e.g., at $10 r_\sun$. 
In the last step in equation~(\ref{eq:ki_cst}) $\zeta=2$ is used. Our discussions assume $\zeta=2$.
We refer $B_0$ and $T_0$ as the footpoint values below. The plasma parcel for the four measurements 
at the footpoints can have very different $n_0$ and $B_0$. {The values of $B_0$ at the footpoints for the four dawn events are found to be $1682.757$ nT, $1054.64$ nT, $966.917$ nT, and $2666.122$ nT for the two Mercury events (nearest Sun first), Venus, and Earth respectively. We find these values by back calculating $|B_{SW}|$ to $10 r_{\sun}$ using the radial scalings discussed above. Because we lack plasma data, we do not know $n_0$ for these four periods.} 
However, if we assume the ratio of $n_{0}/B_{0}$ to have a small variation for the four periods in our study (more on this below), then the $|B|$ dependence from equation$~$(\ref{eq:ki_cst}), $B^{0.5}$, is close but weaker than that in equation~(\ref{eq:kbrk}), which is  $B^{0.61}$.}

{Next if the break is controlled by the Larmor scale $k_{L} =
\Omega_{p}/V_{th}$,  then we can proceed in a similar manner to write,
\begin{equation}
    k_L/C_2 = k_i \beta^{-1/2} =
    \bigg(\frac{B_{0}^{\frac{2}{3\zeta}}}{T_0}\bigg)^{1/2} B^{1-\frac{1}{3\zeta}} 
    \myeq
    \bigg(\frac{B_0^{1/3}}{T_0}\bigg)^{1/2} B^{5/6} ,
    \label{eq:kl_cst}
\end{equation}
where $C_2$ is another constant. Now if we assume the ratio of $ B_0^{1/3}/T_{0} $ to have a small variation for our four measurements, then the $|B|$ dependence from equation~(\ref{eq:kl_cst}), $ B^{0.83} $, is stronger than that in equation~(\ref{eq:kbrk}).}

{If the break is controlled by the cyclotron resonance scale $k_{d} = \Omega_{p}/(V_A+V_{th})$,  then we can proceed  to write,
\begin{equation}
    k_d/C_3 = k_i  \frac{1}{1+\beta^{1/2}}  = k_L  \frac{1}{1+\beta^{-1/2}}  
    \label{eq:kd_cst}
    \end{equation}
where $C_3$ is another constant. The $|B|$ dependence of  
$k_d$ will be between that of $k_i$ and $k_L$ and can therefore be in better agreement with equation~(\ref{eq:kbrk}) than either $k_i$ or $k_L$.}

{Finally, if the break is controlled by the disruption scale as recently proposed by \citep{Mallet2017b, Vech2018, Loureiro2017}, then
\begin{equation}
 k_D/C_4 =   (n_e^{1/2} B/v_{th,e})^{4/9}  = \bigg(\frac{n_{0,e}/T_{0,e}}{B_0^{4/3\zeta}}\bigg)^{2/9}
 B^{4/9+8/27\zeta} \myeq
    \bigg(\frac{n_{0,e}}{T_{0,e} B_0^{2/3}}\bigg)^{2/9} B^{16/27}, 
    \label{eq:kD_cst}
\end{equation}
where $C_4$ is another constant. 
If the prefactor $[n_{0,e}/(T_{0,e} B_0^{2/3})]^{2/9}$ has a small variation 
for our four measurements, then the $|B|$ dependence from equation~(\ref{eq:kD_cst}), $B^{16/27}\sim B^{0.59}$, is almost the same to $\delta_{1}$ in equation~(\ref{eq:fbump}). Note that the exponent $2/9$ of the prefactor in 
equation~(\ref{eq:kD_cst}) is smaller than those in
equations~(\ref{eq:ki_cst}) and (\ref{eq:kl_cst}), therefore the $|B|$ dependence from equation~(\ref{eq:kD_cst}) is less affected by $B_0$, $n_0$, and $T_e$, and is thus more trustworthy than those from equations~(\ref{eq:ki_cst}),
(\ref{eq:kl_cst}), and (\ref{eq:kd_cst}).
}


{Above discussions illustrate 
that if $ln(f_{bump}/k_{brk})$ is independent with $B$ and $r$, then one can obtain certain constraints on the spectral break of the dissipation range. 
By applying radial scalings for $n_p$, $B$, and $T_p$, we can cast the four scales ($k_i$, $k_l$, $k_d$, $k_D$) to a form that has an explicit power law dependence on $B$ and prefactors which involve $n_0$, $B_0$, and $T_0$ at $r=10 r_{\sun}$. Assuming these prefactors do not vary significantly for our events, we find that the disruption scale $k_D$ agrees best with our observation.
The cyclotron resonance scale $k_{d}$ is also 
in general agreements.
The ion inertial scale $k_i$ seems to have a 
shallower $B$ dependence and the gyration scale $k_L$ seems to have a stronger $B$ dependence. Our analysis is consistent and 
supports that of \citep{Vech2018}, who argued that the disruption scale is most likely to be the cause of the dissipation spectral break, although the cyclotron resonance scale is also possible when $\beta_p$ is not extreme \citep{Woodham2018, Duan2020}.}

\section{Summary}
To summarize, in this paper we show that two frequencies: the one which is related to the ``1-Hz" upstream waves at planetary bow shocks, $f_{bump}$, and the one which is related to the solar wind turbulence spectral break, $f_{brk}$,  are strongly correlated.
\mt{We found that the frequencies for the spectral break during 3 planet fly-bys of MESSENGER and one period from MMS are in the range of $0.3$ to $0.7$ Hz, and the frequencies for the spectral bump are in the range of $1.3$ to $2.6$ Hz.} However, the log of their ratio, $\eta=ln f_{bump} - ln f_{brk}$, as measured from the dawn side flybys of MESSENGER, appears to be independent of $|B|$ or $r$.   This weak dependence may be a consequence of a $B$-independent $lnf_{bump} - ln k_{brk}$ where the spreading of $\eta$ reflects the variation of the solar wind speed in these $4$ periods. 
This correlation between $f_{bump}$ and $f_{brk}$ ($k_{brk}$) is unexpected, but connects two seemingly different phenomena, and can be used to discern the underlying mechanisms of the spectral break. {Assuming that $f_{bump}/k_{brk}$ is independent of $|B|$, we discussed its implication on the underlying scale of the dissipation spectral break. We examined four scales: the ion inertial scale $k_i$, the ion gyration scale $k_L$, the cyclotron resonance scale $k_d$, and the disruption scale $k_D$. 
we cast these four scales 
to a form that has an explicit power law dependence on $B$ and prefactors which involve $n_0$, $B_0$, and $T_0$ at $r=10 r_{\sun}$. 
Their $B$ dependence are given by equations~(\ref{eq:ki_cst}), (\ref{eq:kl_cst}), (\ref{eq:kd_cst}), and (\ref{eq:kD_cst}). Among these four, the disruption scale $k_D$ is in the best agreement with our observation. The prefactor of $k_D$ is also the least sensitive to $n_0$, $B_0$, and $T_0$ at the footpoint ($r=10 r_{\sun}$), and therefore yields a $B$ dependence the most trustworthy in all four scales. In comparison, the ion inertial scale predicts a weaker $B$ dependence and the ion gyration scale predicts a stronger $B$ dependence. Because the ion cyclotron scale has a $B$ dependence in between those due to the ion inertial scale
and the ion gyration scale, it is also in general agreement with our observation. Comparing to the disruption scale, it has a stronger dependence on plasma properties through $\beta$ as evident from equation~(\ref{eq:kd_cst}). Note 
that once a thorough understanding of the turbulence spectral break becomes available, our finding of $f_{bump}/f_{brk}$ being independent of $|B|$ will allow one to better understand the underlying generation mechanisms of the upstream waves at planetary bow shocks.}
{To substantiate our claims, further investigations of $f_{bump}/f_{brk}$ from observations where plasma parameters are available are necessary and will be pursued in a future work.} 
\mt{Finally we note that the disruption scale is intimately related to magnetic reconnection.  Behind the Earth's bow shock, magnetic reconnection is abundant in the magnetosheath \citep{Karimabadi2014}.  Consequently one expects that examining turbulence in the magnetosheath using MMS observations, as practiced in, e.g., \citet{Macek2018} can shed lights on the dissipation mechanism. }




\acknowledgments 
MT acknowledges the support of NASA FINESST Fellowship 80NSSC20K1516. This work is supported in part by NASA grants 80NSSC19K0075, and 80NSSC19K0831.



\end{document}